\documentclass[aps,prl,reprint,superscriptaddress]{revtex4-2}

\usepackage{amsmath,amsfonts,amssymb,mathrsfs,xcolor} 
\usepackage{natbib} 
\usepackage{subfigure} 
\usepackage{tabularx} 
\usepackage{epsfig} 
\usepackage{longtable} 
\usepackage{rotating} 
\usepackage{bbold} 
\usepackage{multirow} 
\usepackage{hhline} 
\usepackage{braket} 
\usepackage{setspace} 
\usepackage[normalem]{ulem} 
\usepackage[unicode=true,bookmarks=true,bookmarksnumbered=false,bookmarksopen=false,breaklinks=false,pdfborder={0 0 1},backref=false,colorlinks=true]{hyperref} 
\hypersetup{linkcolor=magenta,urlcolor=blue,citecolor=blue,pdfstartview={FitH},unicode=true}

\def\be{\begin{equation}}
\def\ee{\end{equation}}
\def\bea{\begin{eqnarray}}
\def\eea{\end{eqnarray}}

\newcommand{\Tr}{\operatorname{Tr}}
\newcommand{\im}{\mathrm{i}} 

\begin{document}

\title{Double-exchange ferromagnetism of fermionic atoms in a $p$-orbital hexagonal lattice}

\author{Haoran Sun}
\affiliation{Hefei National Research Center for Physical Sciences at the Microscale and School of Physical Sciences,
University of Science and Technology of China, Hefei 230026, China} 
\author{Erhai Zhao}
\affiliation{Department of Physics and Astronomy, George Mason University, Fairfax, Virginia 22030, USA} 
\author{Youjin Deng}
\affiliation{Hefei National Research Center for Physical Sciences at the Microscale and School of Physical Sciences,
University of Science and Technology of China, Hefei 230026, China}
\author{W. Vincent Liu}
\email{wvliu@pitt.edu}
\affiliation{Department of Physics and Astronomy and IQ Initiative, University of Pittsburgh, Pittsburgh, Pennsylvania 15260, USA} 
\date{\today}

\begin{abstract}
A broad class of correlated quantum materials features strong Hund's coupling, yet cold-atom quantum simulators have so far focused primarily on single-orbital Fermi-Hubbard systems near a Mott insulator. Here we show that repulsively interacting fermions loaded into the $p$-bands of a hexagonal lattice offer a unique platform to study the interplay of ``Hundness'' and ``Mottness.'' Our theory predicts that the orbital degrees of freedom, despite geometric frustration, produce a rich phase diagram featuring a competing itinerant ferromagnetic (FM) metal and a spin-1 antiferromagnetic (AFM) insulator. 
While FM order is known to emerge at low fillings near the flat-band limit, we show that this itinerant ferromagnetism is remarkably robust, persisting to stronger interactions and higher fillings well beyond the flat-band regime.
More importantly, we identify that near half-filling, FM is stabilized by a different mechanism:  the double exchange with spins aligning to avoid Hund-rule penalties at the expense of the kinetic energy of Dirac fermions.
The competition between the kinetic and double-exchange energies leads to, within our approximations, a first order transition between FM and AFM orders in the chemical potential-interaction strength plane.  
Our analysis suggests that $p$-orbital Fermi gases --- as multi-orbital systems with coexisting localized and itinerant spins --- reveal a rich competition between correlated ``Hund metals'' and other unconventional states driven by quantum fluctuations.
  
\end{abstract}

\maketitle 


In many correlated materials~\cite{tokura_orbital_2000} such as cuprates, pnictides, and ruthenates, the $d$- or $f$-orbital degrees of freedom (DOF) play an active role in giving rise to a wealth of cooperative phenomena, ranging from orbital ordering, magnetism, to unconventional superconductivity~\cite{zhang_effective_1988,chen_interplay_2023}. These interacting multi-orbital systems provide a natural arena for the interplay between two paradigms of strong correlation: ``Mottness'', driven by on-site repulsion $U$ which typically leads to antiferromagnetism via superexchange, and ``Hundness'', driven by Hund's rule coupling $J_H$ which promotes high-spin states and itinerant ferromagnetism via mechanisms like double-exchange~\cite{stadler_hundness_2019}. It remains a theoretical challenge to quantify the coexistence and competition of these two mechanisms, in part due to complications from crystal fields, phonons, or spin-orbit coupling. Ultracold atoms in optical lattices can serve as excellent quantum simulators to explore strong correlation physics, e.g., the Fermi-Hubbard model, in a simpler, cleaner, and highly tunable environment~\cite{ibloch_many_2008}. However, so far the experimental efforts have predominantly focused on the lowest $s$-orbital band~\cite{masatoshiimada_metal_1998}, leaving the rich physics inherent in higher orbital bands~\cite{li_physics_2016}, such as in $p$-bands dominated by Hund's coupling, largely unexplored.

Some of the uniquely interesting properties of interacting atoms in the $p$ band of optical lattices have been recognized before. A defining feature of $p$-orbitals is their spatial symmetry which leads to anisotropic hopping patterns distinct from the familiar $e_g$ or $t_{2g}$ orbitals found in many transition metal oxides and, as a result, gives rise to topologically nontrivial band structures in ladders~\cite{li_topological_2013} or flat bands on the hexagonal lattice~\cite{zhang_proposed_2010}. In the Mott limit of strong interaction and integer filling, virtual hopping induces anisotropic kinetic exchange interactions. For example, for spin-polarized $p$-orbital fermions, Refs.~\cite{zhao_orbital_2008,congjunwu_orbital_2008} derived the orbital exchange Hamiltonian to show it is geometrically frustrated on many two-dimensional lattices, and Ref.~\cite{zou_continuum_2016} generalized the 120$^\circ$ model to a tripod model that also contains the Ising and Kitaev models at special limits. For (pseudo)spin 1/2 atoms, the $p$-orbital interaction Hamiltonian (Eq.~\ref{eq:interaction_hamiltonian_full}) derived from the contact $s$-wave interaction between atoms of opposite spins differs from the Hubbard-Kanamori form characteristic of the Coulomb interaction between electrons in solids. For $p$-band fermions on the cubic lattice, it has been noted that the ferromagnetic and antiferromagnetic phases compete and are separated by a paramagnetic state~\cite{wang_magnetism_2008}. There is a large body of work on bosonic atoms on higher orbital bands. For instance, spin-orbital intertwined order and exotic superfluid phases have been predicted for two-component bosons in hexagonal lattices~\cite{li_spin_2021}. 

In this work, we present a comprehensive analysis of interacting fermionic atoms on the $p$-band of hexagonal optical lattice, where the coexistence of flat band and dispersive Dirac cones, multi-orbital interactions, and Fermi statistics conspire to provide a fertile ground for competing orders. Previous studies have identified flat-band ferromagnetism (FM) at ultra low fillings \cite{wu_flat_2007} and a staggered antiferromagnetic (AFM) insulator \cite{zhang_proposed_2010} at half-filling. The physics beyond these two special limits has remained inconclusive and mysterious. It was conjectured that phase coexistence, orbital order, or even ferrimagnetic order are possible~\cite{zhang_proposed_2010}. Our work is motivated to address the following open questions: What is the mechanism that stabilizes ferromagnetism when doping away from the half-filled AFM insulator? And what is the nature of the transition, if any, between the AFM insulator and FM metal?

Our analysis yields a few unexpected results. First, long-range orbital order is suppressed by geometric frustration on the hexagonal lattice, in sharp contrast to what happens on the square lattice~\cite{zhou_spin_2015,zhao_orbital_2008}, so that the phase diagram is dominated by the competition between distinct spin orders. Second, the itinerant FM phase is robust and extends to higher fillings and strong interactions. The flat band provides a solid foundation to promote FM, but it is not exactly a prerequisite. Third, upon doping the AFM insulator, the FM phase remains stabilized by the celebrated double-exchange mechanism. Within mean-field theory, the competition between FM and AFM near half filling produces a density-discontinuous region consistent with phase separation. Throughout the FM phase, itinerant carriers coexist with localized moments, even though the underlying physics responsible for the localization is different in two limits: the localized states are associated with the flat band at low filling, while spin-1 moments are formed due to ``Mottness'' near half filling. While all terms in the interaction Hamiltonian contribute, as we will explain below, the persistence of FM is largely due to Hund's coupling.

Realizing the double-exchange mechanism, a cornerstone of magnetism in solids~\cite{zener_interaction_1951,anderson_considerations_1955,koch_exchange_2012}, has been a long-standing challenge for quantum simulators~\cite{bugnion_exploring_2013}. It emerges naturally in our setup, which therefore provides a clean cold-atom route to double exchange and Hund physics in a multi-orbital setting, with the FM-AFM boundary near half filling serving mainly as a signature of their competition. Our results pinpoint the relevance of ``Hundness'' in orbital quantum gases and provide concrete theoretical predictions for available higher-band optical-lattice experiments~\cite{wirth_evidence_2011,soltanpanahi_quantum_2012,hachmann_quantum_2021,kiefer_ultracold_2023,jin_manipulation_2022}.

\begin{figure}[!htp]
    \centering
    \includegraphics[width=\columnwidth]{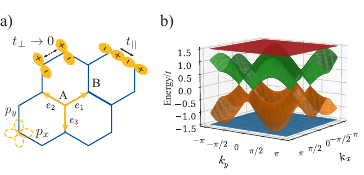}
    \caption{(a) Hexagonal lattice with A/B sublattices and nearest-neighbor vectors; transverse hopping $t_\perp$ is neglected. (b) Non-interacting bands for $t_\perp = 0$, with two inequivalent Dirac cones and two perfectly flat bands at $\epsilon = \pm 3t/2$.} 
    \label{fig:hexagonal_lattice_bandstructure}
\end{figure}

\textit{Model Hamiltonian}. We consider spin-1/2 fermions loaded in the $p$-bands of a hexagonal optical lattice, assuming the lower $s$ band is completely filled and well separated from the $p$-bands. The system's kinetics are described by a tight-binding Hamiltonian with nearest-neighbor hopping:
\begin{equation}
\label{eq:hopping_hamiltonian}
H_t = \sum_{\vec{r} \in A, i, \sigma}  t_{\parallel} c_{i \sigma}^{\dagger}(\vec{r}) c_{i \sigma}(\vec{r}+\hat{e}_i) + \text{H.c.}
\end{equation}
Here, $c_{i \sigma}(\vec{r})$ annihilates a fermion with spin $\sigma$ at site $\vec{r}$ with the $A$ sublattice in an orbital projected to the direction of unit vector $\hat{e}_i$ (as a superposition of $p_x$ and $p_y$), $t_\parallel$ is the longitudinal hopping amplitude (for details see Appendix~\ref{app:orbital_projection}). In typical experiments, transverse hopping is negligible, $t_\perp \ll t_\parallel$, and we set $t_\parallel \equiv t = 1$ as our energy unit. The validity of this approximation in $H_t$ and the restriction to the $p$-orbital manifold rests on a clear hierarchy of energy scales found in typical optical lattice experiments, as detailed in Appendix~\ref{app:energy_scales}. The resulting non-interacting band structure features two perfectly flat bands touching two dispersive bands with Dirac cones as shown in Fig.~\ref{fig:hexagonal_lattice_bandstructure}.

The interaction Hamiltonian is obtained by approximating the local optical potential, assumed deep, as a harmonic oscillator and decomposing the on-site $s$-wave repulsion in the Wannier basis, keeping only the $p$-orbital manifold~\cite{liu_atomic_2006,aisacsson_multi_2005}:
\begin{align}
\label{eq:interaction_hamiltonian_full}
H_{\text{int}} = & \frac{3U}{4}\left(n_{x\uparrow}n_{x\downarrow} + n_{y\uparrow}n_{y\downarrow}\right) 
+ \frac{U}{4}\left(n_{x\uparrow}n_{y\downarrow} + n_{y\uparrow}n_{x\downarrow}\right)  \nonumber\\ 
&+ \frac{U}{4}\left(\Delta_x^\dagger\Delta_y + \Delta_y^\dagger\Delta_x  - S_x^+S_y^- - S_y^+S_x^- \right)
\end{align}
where $n_{\mu\sigma}=c_{\mu\sigma}^\dagger c_{\mu\sigma}$ is the number operator, $\Delta_\mu = c_{\mu\downarrow} c_{\mu\uparrow}$ is the intra-orbital pair transfer operator, and $S_\mu^+ = c_{\mu\uparrow}^\dagger c_{\mu\downarrow}$ is the spin raising operator for orbital $\mu=x,y$. The interaction energy scale $U>0$ is the on-site interaction for atoms in the $s$-orbital band. The Wannier functions for $p$-orbitals have smaller overlap, so the intra-orbital repulsion is $3U/4$~\cite{liu_atomic_2006,aisacsson_multi_2005}. The remaining terms are the inter-orbital repulsion, pair transfer, and Hund's coupling (notice the sign), $- \frac{U}{4}(S_x^+S_y^- + S_y^+S_x^-)$, that favors spin alignment on the same site. The repulsion terms are foundational to Mott physics~\cite{masatoshiimada_metal_1998}, while the Hund's coupling is the primary driver of ferromagnetic spin correlations~\cite{georges_strong_2013, stadler_hundness_2019}. This operator structure is generic for multi-orbital systems and is also found for Coulomb interactions, albeit with different prefactors~\cite{pfazekas_lecture_1999}.

We can rewrite $H_{\text{int}}$ so that the competing orders become more transparent. Define the pseudo-spin operators $T^j_\sigma = \frac{1}{2} \sum_{\mu,\mu'} c_{\mu\sigma}^\dagger \tau^j_{\mu\mu'} c_{\mu'\sigma}$ (a vector in orbital space for each spin $\sigma$) and the spin operators $S^j_\mu = \frac{1}{2} \sum_{\sigma,\sigma'} c_{\mu\sigma}^\dagger \sigma^j_{\sigma\sigma'} c_{\mu\sigma'}$ (for each orbital $\mu$), then~\cite{zhou_spin_2015}
\begin{equation}
\label{eq:interaction_S_operators}
H_{\text{int}} = -\frac{U}{2} \left[ (T^y_\uparrow + T^y_\downarrow)^2 + (\vec{S}_x + \vec{S}_y)^2 \right] + \frac{U}{2}n.
\end{equation}
where $n=\sum_{\mu\sigma} n_{\mu\sigma}$ is the total number operator. 

This form decomposes the interaction into spin and orbital channels, making the SU(2) spin symmetry and SO(2) orbital rotation symmetry manifest.  The orbital SO(2) generator is $T^y\equiv \sum_\sigma T^y_\sigma$, which plays a dual role: it is both a component of the pseudospin vector and the on-site orbital angular momentum $L^z = 2T^y= -i \sum_{\sigma} (c_{x\sigma}^\dagger c_{y\sigma} - c_{y\sigma}^\dagger c_{x\sigma})$. Under the orbital transformation, $(c_{x\sigma},c_{y\sigma})^T \to (c_{x\sigma}',c_{y\sigma}')^T = e^{-i\varphi\tau^y}(c_{x\sigma},c_{y\sigma})^T$, the operator $T^y=L^z/2$ remains invariant. This ensures that $H_\mathrm{int}$ is formulated in a basis that naturally preserves $L^z$ angular momentum, rather than artificially imposing a preferred orbital axis.
This symmetry-explicit representation is particularly amenable to field-theoretic analysis: the squared operators can be decoupled by introducing auxiliary fields---a scalar field for the orbital channel and a vector field for the spin channel---via Hubbard-Stratonovich transformations~\cite{koinov_a_2010}.

\begin{figure}[!htp]
    \centering
    \includegraphics[width=\columnwidth]{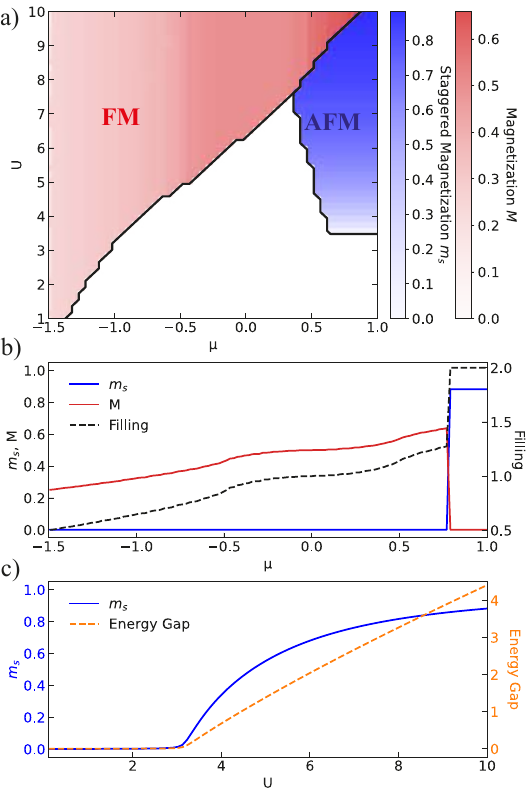}

    \caption{
        Overview of spin order and phase competition.
        (a) Mean-field phase diagram in the $\mu$-$U$ plane (units: $t\equiv 1$). The color map shows the order-parameter magnitude of the FM phase (total magnetization $M$, red) and AFM phase (staggered magnetization $m_s$, blue); the white region is paramagnetic within mean-field.
        (b) Strong-interaction slice at $U=10$, showing $m_s$, $M$, and $\langle n \rangle$ versus $\mu$. Discontinuous jumps in all three quantities signal a first-order FM-AFM boundary, with density discontinuity consistent with phase separation.
        (c) Half-filled slice ($\langle n \rangle = 2$). The staggered magnetization $m_s$ and gap versus $U$ show a continuous AFM transition at $U_c \approx 3.253t$.
    }
    
    \label{fig:main_phase_diagram}
\end{figure}

{\it Absence of orbital order}. The full Hamiltonian $H_t + H_{int}$ for general values of $U$ and filling fraction of the $p$-bands presents a formidable many-body problem. In this work, we limit our attention to a simple, yet unbiased and comprehensive, approximation scheme. As detailed in Appendix, we first express $H_{int}$ in terms of (spin- and site-resolved) orbital pseudo-spin operators $T^j_{i,\sigma}$ and define order parameter $t^j_{i,\sigma} = \langle T^j_{i,\sigma} \rangle$, then decouple $H_{int}$ into a quadratic mean-field Hamiltonian $H_{MF}$ and solve $H_t + H_{MF}$ self-consistently with the number equation at zero temperature. We allow all possible spin and orbital wave configurations and do not assume a priori the (co-)existence or absence of any particular order. The resulting ground-state phase diagram in the chemical potential ($\mu$) versus interaction ($U$) plane is presented in Fig.~\ref{fig:main_phase_diagram}. Previous works only discussed certain limits of this global phase diagram, and our calculation serves as the prerequisite step toward understanding the intricate phases in this multi-orbital system.

\begin{figure}[!htp]
    \centering
    \includegraphics[width=0.9\columnwidth]{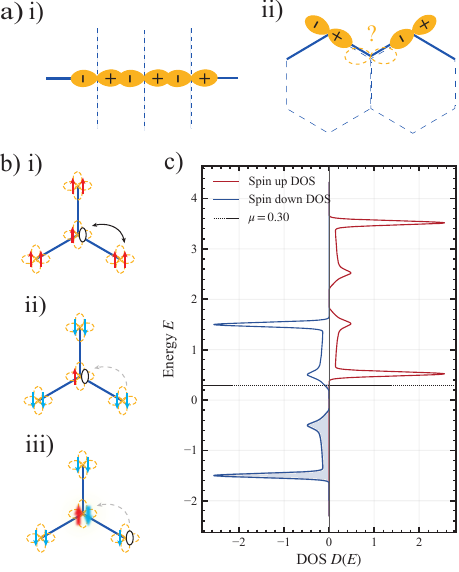}
    \caption{Schematics of competing mechanisms governing the ground state.
    (a) Geometric orbital frustration on the hexagonal lattice. Unlike the unfrustrated ferro-orbital alignment on a square lattice (left), the hexagonal geometry prevents simultaneous kinetic-energy minimization on all bonds due to the anisotropic $p$-orbital structure (right), suppressing long-range orbital order.
    (b) Double exchange mechanism. (i) A FM background lets holes delocalize, whereas hopping in (ii-iii) an AFM background is blocked by the on-site Hund penalty $J_H=U/4$ (blurred spins), favoring FM at finite doping.
    (c) Spin-resolved DOS at $U = 8, \mu=0.3$ normalized per lattice site. Pronounced flat-band peaks indicate localized moments, while the low-energy carriers are dominated by a single spin species. The flat-band contribution is artificially broadened for visualization to avoid singular behavior. At this parameter configuration, the Stoner criterion $U D(E_F)=0.73<1$ is not satisfied.}
    \label{fig:orbital_frustration}
\end{figure}

The absence of long-range orbital order in Fig.~\ref{fig:main_phase_diagram} results from the competition between spin and orbital sectors inherent in the interaction Eq.~\ref{eq:interaction_S_operators}. While the bipartite hexagonal lattice supports unfrustrated magnetic orders, the orbital channel is suppressed by strong geometric frustration. Unlike on the square lattice~\cite{zhou_spin_2015}, the hexagonal geometry precludes simultaneous bond-energy minimization for orbitals (Fig.~\ref{fig:orbital_frustration}). Consequently, the frustrated orbital channel is energetically costly, ceding the ground state to the unfrustrated spin channel. This mechanism mirrors the frustration observed in $p$-orbital bosons~\cite{li_spin_2021} and spinless fermions~\cite{congjunwu_orbital_2008,zhao_orbital_2008} on hexagonal lattice, where complex circulating orbital textures are predicted.

The stability against orbital ordering can be further confirmed by examining the collective orbital excitations over a postulated (e.g. FM or AFM) state. A field-theoretic calculation confirms that the orbital excitations are fully gapped throughout the phase diagram (see Appendix~\ref{app:orbital_excitation} for details). The finite gap ensures that the system is not close to any instability toward orbital order. Consequently, the unfrustrated spin order prevails.

{\it Kinetic exchange for spin-1}.
With orbital order suppressed, the phase diagram consists of a paramagnetic (PM) metal, a ferromagnetic (FM) metal, and a half-filled antiferromagnetic (AFM) insulator. Here the PM label should be understood within the present mean-field ansatz; once fluctuations beyond mean field are included, this region may instead be a correlated metal. At half-filling ($\langle n \rangle = 2$), for interactions $U$ above a critical value $U_c$, the system undergoes a second-order transition from a semi-metal to an AFM insulator. This AFM phase is characterized by a non-zero staggered magnetization, $m_s$, which modulates the local density according to: 
\begin{equation}
\label{eq:staggered_magnetization_order_parameter}
    \langle n_{\alpha, \sigma} \rangle = 1 + \alpha \sigma m_s.
\end{equation}
Here, $\alpha = \pm 1$ for sublattices A/B and $\sigma = \pm 1$ for spin up/down, and the orbital index $\mu$ is implicitly summed over due to the absence of orbital order. Opening a Mott gap at the Dirac points is described by
\begin{equation}
\label{eq:gap_equation}
    \frac{4}{U} = \frac{1}{\sqrt{{9}/{4} + (Um_s)^2}} + \int d\epsilon \, \frac{g(\epsilon)}{\sqrt{\epsilon^2 + (Um_s)^2}},
\end{equation}
where $g(\epsilon)$ is the density of states of the Dirac bands. This yields a critical interaction strength of $U_c \approx 3.253t$.

As a signature of the interacting $p$-orbital system, this AFM phase differs from the familiar single-band Hubbard model. At half filling, each lattice site is occupied by two fermions, and strong on-site interactions $H_{int}$ (specifically, the Hund's coupling) align their spins to form a local spin-1 moment~\cite{zhang_proposed_2010}. To see the AFM order, consider virtual hopping processes induced by perturbation $H_t$. The dominant contribution comes from two parallel spins hopping to a neighboring site populated by two opposite spins, yielding an intermediate state with four spins on the same site before hopping back. The picture is similar to the standard kinetic exchange, but for spin-1 moments. Perturbation theory shows that the effective interaction $J_{AFM}$ is antiferromagnetic and its magnitude scales as $J_{AFM} \sim 16t^4/U^3$. In this sense, it is a weaker antiferromagnet. As we shall see below, interesting things happen when it is doped with holes.

\textit{Beyond flat-band ferromagnetism}. The emergence of ferromagnetic (FM) order at low fillings, $0.25<n<0.5$, and small $U$ is triggered by the flat band inherent to $p$-orbitals on the hexagonal lattice~\cite{wu_flat_2007}. The flat band supports a set of degenerate, compact localized (``plaquette'') states, and the FM order results from a direct exchange mechanism mediated by virtual hopping between localized spin-1/2 moments~\cite{zhang_proposed_2010}. The on-site interaction Hamiltonian (Eq.~\ref{eq:interaction_hamiltonian_full}) explicitly includes a Hund's rule coupling term, making the effective spin-spin interaction favor a spin-aligned state. This result is also in accordance with the Mielke--Tasaki theorem for flat-band ferromagnetism in Hubbard-like systems~\cite{mielke_ferromagnetism_1993,pons_flat_2020}.

For filling $n>0.5$ and finite $U$, however, it is no longer sufficient to consider only the plaquette states or the flat band, because particles can be lifted by the interaction from the flat band into the dispersive lower Dirac band. Despite this, at small $U$ and low filling, it is still useful to view the system as self-organized into two parts: mobile carriers plus a ferromagnetic background of localized spins. Orbital-induced Hund's coupling prefers a mobile carrier to align its spin with the FM background. In other words, the interaction $U$ stabilizes and promotes ferromagnetism. On the other hand, placing mobile carriers in higher bands costs kinetic energy relative to keeping opposite-spin particles in lower-energy states. 

Thus, for fixed $U$ and increasing $\mu$, as more and more carriers are added, the FM order will eventually become disfavored and give way to a paramagnetic metal (PM). Our calculation confirms this intuition and predicts a FM-PM phase boundary extending to large $U$ and $\mu$. The FM phase is characterized by its order parameter, the magnetization $M$, 
\begin{equation}
\label{eq:ferromagnetic_order_parameter}
    M = \frac{1}{2N} \sum_{\vec{r}} \langle n_{\vec{r},\uparrow} \rangle - \langle n_{\vec{r},\downarrow} \rangle,
\end{equation}
where $N$ is the total number of sites. In Fig.~\ref{fig:main_phase_diagram}(a) and (b), $M$ only depends weakly on $U$ but increases with $\mu$. This agrees with the qualitative picture above that added carriers contribute to the magnetization. The resulting FM regime occupies a broad region of the phase diagram, showing that the underlying mechanism is not restricted to ultra-low filling or an asymptotic strong-coupling limit. The robustness of FM order, well beyond the flat band regime studied before, is largely due to Hund's coupling.

{\it Double-exchange-driven AFM-FM transition.} 
In the unexplored regime near half filling and strong $U$, the mechanism of flat-band ferromagnetism becomes irrelevant. To understand how hole doping drives an FM instability, consider a single hole introduced into the half-filled spin-1 AFM insulator. The remaining particle on the doped site ($\uparrow \circ$) occupies the flat band; consequently, it acts as a localized magnetic moment, mirroring the core electrons in transition metal magnets. As illustrated in Fig.~\ref{fig:orbital_frustration}b(ii) and (iii), carrier transport in an AFM background is energetically suppressed: if a particle hops onto a hole-doped site ($\uparrow\circ $) from a neighbor with antiparallel spin ($\downarrow\downarrow$), the resulting doubly occupied site forms a low-spin ($S = 0$) state, violating Hund's rule and incurring an energy penalty $J_H = U/4$. As a result, the effective hopping of the hole is suppressed. In contrast, a ferromagnetic background allows the itinerant carrier to hop to neighboring sites while maintaining the energetically favorable high-spin ($S=1$) configuration, as shown in Fig.~\ref{fig:orbital_frustration}b(i). This allows the holes to fully delocalize and maximize their kinetic energy gain. When this kinetic gain exceeds the cost of breaking the antiferromagnetic superexchange bonds, the itinerant carriers enforce ferromagnetic alignment on localized spins to optimize their mobility. This process realizes the double-exchange mechanism~\cite{zener_interaction_1951,anderson_considerations_1955,koch_exchange_2012}, the fundamental driver of ferromagnetism in conductive transition metal oxides such as manganites and magnetite. 

The physical picture outlined above is supported by our numerics on the spin-resolved DOS shown in Fig.~\ref{fig:orbital_frustration}(c): pronounced flat-band peaks indicate localized moments, while the carriers near the Fermi level are dominated by a single spin species aligned with them. Moreover, the corresponding Stoner quantity is only $UD(E_F)=0.73<1$, so the FM state cannot be attributed to a simple Stoner instability.

As a secondary consequence of this competition, the mean-field solution near the FM-AFM boundary exhibits a density-discontinuous region consistent with phase separation. Its onset is controlled by the competition between the kinetic energy gain from hole delocalization in the FM state, $\Delta E_{FM}$, and the interaction energy gain from antiferromagnetic order, $\Delta E_{AFM}\propto J_{AFM}(1-\delta)^2$. For small hole doping $\delta = 2 - \langle n \rangle$, the kinetic energy of the metallic FM phase can be computed from the linear density of states of the Dirac bands, $\Delta E_{FM} \propto t\delta^{3/2}$. Equating $\Delta E_{FM}$ to $\Delta E_{AFM}$ yields the scaling of the AFM-FM phase boundary in terms of critical doping $\delta_c \propto (t/U)^2$. This smaller scale controls the local destabilization of the AFM state at the onset of phase separation, and the trend agrees with the self-consistent mean-field result in Fig.~\ref{fig:main_phase_diagram}. In particular, as $U\rightarrow \infty$, the AFM becomes unstable at infinitesimal hole doping.

We emphasize that the landscape of competing magnetic orders shown in Fig.~\ref{fig:main_phase_diagram} is a direct consequence of the orbital degree of freedom. The standard $s$-band Hubbard model on the hexagonal lattice does not host a stable itinerant FM phase; it only exhibits a transition from a paramagnetic metal to an AFM insulator~\cite{raczkowski_the_2020, ostmeyer_the_2020, zeng_phase_2022}. It is interesting that in our system, depending on the filling and interaction strength, three different mechanisms are at play: flat-band ferromagnetism, kinetic exchange for spin-1 AFM, and double-exchange. Throughout, Hund's coupling that favors spin-aligned configurations plays an important role.

We establish that interacting fermions in the $p$-orbitals of a hexagonal lattice provide not only a natural platform for quantum simulating double exchange, a cornerstone of magnetism in solids~\cite{zener_interaction_1951,anderson_considerations_1955}, but also access to a previously unexplored parameter regime beyond standard condensed-matter settings. In our system, localized moments and itinerant carriers emerge intrinsically from the flat and dispersive Dirac bands, thereby overcoming the long-standing challenge of engineering double exchange with cold atoms~\cite{maciejlewenstein_ultracold_2007}. The orbital degree of freedom further generates pronounced Hund physics, yielding a robust ferromagnetic phase far beyond the flat-band limit through a distinct double-exchange mechanism. Near half filling, the competition between FM and AFM also produces a density-discontinuous region consistent with phase separation, which we interpret as supporting evidence for that mechanism. Although geometric frustration suppresses long-range orbital order, the orbitals remain essential; unlike in the single-band Hubbard model, Hund's coupling in Eq.~\ref{eq:interaction_hamiltonian_full} reshapes the spin sector by stabilizing itinerant ferromagnetism at low fillings and forming local spin-1 units that underpin the antiferromagnetic state at half filling~\cite{georges_strong_2013}.

These predictions are experimentally testable with available higher-band optical-lattice techniques~\cite{ibloch_many_2008,wirth_evidence_2011,soltanpanahi_quantum_2012,hachmann_quantum_2021,kiefer_ultracold_2023,jin_manipulation_2022}. The distinct phases, from staggered order to double-exchange-driven ferromagnetism, can be identified with time-of-flight imaging, momentum-resolved spectroscopy, and in situ microscopy~\cite{bakr_a_2009,sherson_single_2010}. The experimentally most relevant target is the robust FM regime itself: its characteristic energy scale is at least competitive with established fermionic optical-lattice experiments. By contrast, resolving the fine structure of the phase competition near the first-order boundary is more demanding because the onset of the density-discontinuous region is controlled by a much smaller energy scale. This study lays the groundwork for incorporating quantum fluctuations using numerical methods such as DMFT or QMC~\cite{georges_dynamical_1996,vanhoucke_feynman_2012}. It may shed further light on the paramagnetic metal, which is consistent with a Hund metal within mean-field theory and may develop other many-body instabilities such as unconventional superfluidity.

\textit{Acknowledgments}. The authors gratefully acknowledge helpful discussions with Yihan Duan, Andreas Hemmerich, Xiaopeng Li, and Peter Zoller. This work is supported by the National Natural
Science Foundation of China (NSFC) under Grant No.
12204173 and No. 12275263, the Innovation
Program for Quantum Science and Technology under
Grant No. 2021ZD0301900 (H.S. and Y.D.), and the AFOSR Grant No.~FA9550-23-1-0598 (E.Z. and W.V.L.).

\bibliographystyle{apsrev4-2}

\bibliography{paper_clean} 

\appendix
\renewcommand{\theequation}{A\arabic{equation}}
\renewcommand{\thefigure}{A\arabic{figure}}
\setcounter{equation}{0}
\setcounter{figure}{0}

\section{Orbital Projection along Hopping Directions}
\label{app:orbital_projection}

The kinetic term in the Hamiltonian, Eq.~(\ref{eq:hopping_hamiltonian}), is expressed in a basis that rotates with the lattice bond direction. This is achieved by projecting the operators for the site-fixed Cartesian orbitals, $\{c_x, c_y\}$, onto axes parallel and perpendicular to the bond vectors.

The hopping term connects sites on sublattice A with neighboring sites on sublattice B. The unit vectors for the longitudinal hopping ($t_\parallel$) term are:
\begin{equation}
    \hat{e}_{1,2} = \pm \frac{\sqrt{3}}{2} \hat{x} + \frac{1}{2} \hat{y}, \quad \hat{e}_3 = -\hat{y}.
\end{equation}
The operators $c_{\vec{r}, i, \sigma}$ annihilate a fermion in an orbital oriented along the corresponding bond direction $\hat{e}_i$. They are defined as the projection of the Cartesian orbital operators:
\begin{equation}
    c_i = (\hat{c}_x \hat{x} + \hat{c}_y \hat{y}) \cdot \hat{e}_i.
\end{equation}
where $\hat{x},\hat{y}$ are unit vectors in corresponding directions. Similarly, the neglected transverse hopping term 

\begin{equation}
   H_{\perp} = \sum_{\vec{r} \in A, i, \sigma}  t_{\perp} \tilde{c}_{i \sigma}^{\dagger}(\vec{r}) \tilde{c}_{i \sigma}(\vec{r}+\hat{e}_i)
\end{equation}
is formulated by operator $\tilde{c}_{i\sigma}(\vec{r})$ corresponds to an orbital oriented perpendicular to the bond direction. The rotated unit vectors, $\vec{e}'_i$, can be obtained by a $\pi/2$ counter-clockwise rotation of the unit vectors, $\hat{e}'_i = R(\pi/2) \vec{e}_i$. This gives
$
    \hat{e}'_{1,2} = -\frac{1}{2} \hat{x} \pm \frac{\sqrt{3}}{2} \hat{y}, \quad \hat{e}'_3 = \hat{x}.
$
The transverse hopping operators are the projection onto these rotated directions. This formulation captures the anisotropic nature of the $p$-orbital wavefunctions, where the hopping integral's value is determined by the relative orientation of the orbitals and the bond connecting them.

The single-particle spectrum derived from the free Hamiltonian with $t_\perp=0$ consists of two flat bands at energies $\epsilon_{\text{flat}} = \pm \frac{3}{2}$ and two dispersive Dirac bands. The dispersion of the latter is given by
\begin{equation}
\epsilon(\vec{k}) = \pm \left[ \frac{3}{2} + \cos(\sqrt{3}k_x) + 2\cos\left(\frac{1}{2}k_y\right) \cos\left(\frac{\sqrt{3}}{2}k_x\right) \right]^{1/2},
\end{equation}
which are symmetric about the zero-energy plane, as illustrated in Fig.~\ref{fig:hexagonal_lattice_bandstructure}(b).

\section{Experimental Realization and Energy Scales}
\label{app:energy_scales}

The validity of our theoretical framework, which is confined to the $p$-orbital manifold and neglects transverse hopping ($t_\perp = 0$), rests on a specific hierarchy of energy scales that is naturally realized in ultracold atom experiments~\cite{wirth_evidence_2011,soltanpanahi_quantum_2012,hachmann_quantum_2021,kiefer_ultracold_2023}. This hierarchy is given by
\begin{equation}
 \Delta_{sp} \gg U \sim t_{\parallel} \gg t_{\perp} \sim t_s,
\end{equation}
where $\Delta_{sp}$ is the energy gap between the $s$- and $p$-orbital bands and $t_s$ is the $s$-orbital hopping. These energy scales are tunable functions of the optical lattice depth $V$, usually expressed in units of the recoil energy $E_R = \hbar^2 k_L^2 / (2m)$, where $k_L=2\pi/\lambda$ is the laser wavevector. In the deep lattice limit, the $s$-$p$ band gap arises from the harmonic confinement at the bottom of each potential well, scaling as $\Delta_{sp} \propto \sqrt{V E_R}$. In contrast, the tunneling amplitudes are exponentially suppressed, with $t_s \propto (V/E_R)^{3/4} \exp(-2\sqrt{V/E_R})$~\cite{djaksch_cold_1998}. A key feature of $p$-orbital systems is that the transverse hopping $t_{\perp}$ has the same functional dependence and magnitude as the $s$-orbital hopping $t_s$, while the longitudinal hopping is significantly larger, scaling roughly as $t_{\parallel} \propto (\frac{\pi^2}{2}\sqrt{V/E_R}) t_s$~\cite{aisacsson_multi_2005}. The interaction strength $U$ depends on the $s$-wave scattering length $a_s$ and scales as $U \propto (k_L a_s) (V/E_R)^{3/4}$~\cite{djaksch_cold_1998}.

To provide a concrete example, a typical, experimentally realistic setup with $^{40}$K atoms at a lattice depth of $V=12 E_R$ and a tuned scattering length $a_s = 500 a_0$ yields the ratios $U/t_{\parallel} \approx 9.4$ and $t_{\parallel}/t_{\perp} \approx 12$. Crucially, the $s$-$p$ gap remains the largest scale, with $\Delta_{sp}/U \approx 4.3$. This demonstrates that even in the strongly correlated regime where interactions $U$ dominate over the kinetic energy $t_{\parallel}$, the system is well protected from hybridization with the $s$-orbital band, firmly validating our $p$-orbital-only model.

As additional experimental benchmarks, AFM correlations, long-range AFM order, and the 3D AFM phase transition in $s$-band Fermi-Hubbard systems have already been observed~\cite{hart_observation_2015,parsons_site-resolved_2016,mazurenko_cold-atom_2017,shao_antiferromagnetic_2024}. While the AFM superexchange in our $p$-band system is a fourth-order process ($16t_\parallel^4/U^3$), the longitudinal $p$-band hopping $t_\parallel$ is significantly larger than the $s$-band hopping $t_s$. Furthermore, as highlighted in the main text, one does not need to resolve the extremely small energy difference of this fourth-order process to observe the robust double-exchange ferromagnetic phase, which is separated from the PM phase by a much larger energy scale. Together with established higher-band fermionic preparations showing long lifetimes~\cite{hachmann_quantum_2021,kiefer_ultracold_2023}, these results support the experimental feasibility of exploring this phase competition.

For exploring strongly correlated magnetic phases, our relevant parameter regime does not demand divergent interactions. Consequently, strong Feshbach-resonant enhancement is not a prerequisite; the target regime is accessible through a combination of natural background repulsion and precise lattice-depth control, with moderate interaction tuning serving as an additional degree of freedom. This approach is particularly advantageous for experimental stability, as operating away from resonance helps suppress three-body loss. Furthermore, in contrast to $p$-band bosonic systems---which have already demonstrated sufficient metastability for systematic study---three-body loss in $p$-band Fermi gases is expected to be reduced by several orders of magnitude due to Pauli blocking~\cite{hachmann_quantum_2021,kiefer_ultracold_2023}.

Regarding state preparation, two potential implementations are feasible for the effective $p$-band model considered in this work. One route uses a filled $s$ band together with partial $p$-band population. The other follows the standard Hamburg-style higher-band protocol, where the $s$ band is left empty and atoms are prepared in a metastable $p$-band state~\cite{wirth_evidence_2011,hachmann_quantum_2021,kiefer_ultracold_2023}. Since our theory is formulated as an effective $p$-band Hamiltonian, either preparation route maps to the same low-energy description used in the manuscript.

\section{Stability Against Orbital Ordering}
\label{app:orbital_excitation}

A central claim in the main text is that the hexagonal lattice geometry frustrates and suppresses long-range orbital order. To substantiate this, we perform a field-theoretic analysis of the collective orbital excitations. We show that these excitations remain gapped throughout the physically relevant phases, indicating stability against an orbital ordering instability.

We begin with the on-site interaction Hamiltonian, expressed in terms of spin and orbital operators as given in Eq.~(\ref{eq:interaction_S_operators}) of the main text. This form is particularly amenable to a Hubbard-Stratonovich decoupling. We introduce auxiliary bosonic fields: a scalar field $\theta(\tau, \vec{r})$ coupling to the orbital channel $(T^y_\uparrow + T^y_\downarrow)$, and a vector field $\vec{\phi}(\tau, \vec{r})$ coupling to the spin channel $(\vec{S}_x + \vec{S}_y)$. Dropping the constant term $\frac{U}{2}n$, the effective action is:
\begin{equation}
\label{eq:app_effective_action}
    S_{\text{eff}}[\vec{\phi}, \theta] = -\int d\tau \left( \frac{1}{2}\theta^2 + \frac{1}{2}|\vec{\phi}|^2\right) - \Tr \ln G^{-1}[\vec{\phi}, \theta],
\end{equation}
where $G^{-1} = \partial_\tau - H_0 - M(\vec{\phi}, \theta)$ is the inverse Green's function for the fermions. $H_0$ is the non-interacting Hamiltonian and $M$ represents the mean-field coupling.

We analyze the stability of the system by examining the fluctuations around the magnetic saddle point. We expand the action around the configuration determined by the self-consistent mean-field solution:
\begin{equation}
    \vec{\phi}_c = \phi_z(-1)^\alpha \hat{z}, \quad \quad \theta_c = 0.
\end{equation}
Here, $\alpha$ is the sublattice index. The vanishing orbital expectation value $\theta_c=0$ is the result of the mean-field energy minimization, consistent with the suppression of orbital order.

The inverse Green's function for the fermions in this background $\vec{\phi}_c$, written in the sublattice basis $\Psi(\vec{k}) = (c_{\mu\sigma A}(\vec{k}), c_{\mu\sigma B}(\vec{k}))^T$, is:
\begin{equation}
\label{eq:app_inverse_greens_function}
G_c^{-1}(k) = \omega - H_t(\vec{k}) \otimes I_{\text{spin}} - \phi_z I_{\text{orb}} \otimes \sigma^z_{\text{spin}} \otimes \sigma^z_{\text{site}}.
\end{equation}

Expanding the action to second order in the orbital fluctuations $\theta(\vec{p}, \Omega)$ (where $\theta$ now denotes the fluctuation field around $\theta_c = 0$) yields:
\begin{equation}
\label{eq:app_action_quadratic_fluctuation}
S^{(2)}_\theta[\theta] = \frac{1}{2} \int \frac{d^2p}{(2\pi)^2} \frac{d\Omega}{2\pi} |\theta(\vec{p},\Omega)|^2 K_\theta(\vec{p},\Omega;\phi_z),
\end{equation}
where $K_\theta(\vec{p},\Omega;\phi_z)$ is the inverse orbital propagator. Physically, it quantifies the energy stiffness against orbital fluctuations. A magnetic phase is stable against orbital ordering if the static stiffness $K_\theta(\vec{p}\to 0, \Omega \to 0)$ is strictly positive (indicating a massive, gapped mode), while a vanishing kernel would signal a divergence in susceptibility and an instability toward an ordered state.

In the static, long-wavelength limit ($\vec{p} \to 0, \Omega = 0$), the kernel takes the standard one-loop form:
\begin{equation}
\label{eq:app_kernel_zero_p_omega}
K_\theta(0,0;\phi_z) = 1 - \frac{\im U}{4} \int \frac{d^2k}{(2\pi)^2} \int \frac{d\omega}{2\pi} f(\vec{k},\omega;\phi_z).
\end{equation}
Here, the integral term represents the static polarization bubble (particle-hole susceptibility), with $f(\vec{k},\omega;\phi_z)$ defined from the trace over the fermion Green's function loop:
\begin{equation*}
f(\vec{k},\omega;\phi_z) = \Tr \left[ G_c(\vec{k},\omega;\phi_z) \Gamma_{\theta} G_c(\vec{k},\omega;\phi_z) \Gamma_{\theta} \right],
\end{equation*}
where the vertex $\Gamma_{\theta} = \tau^y_{\text{orb}} \otimes I_{\text{spin}} \otimes I_{\text{site}}$ describes the coupling of the orbital field to the fermions. 

For the semimetallic phase where $\phi_z = 0$, numerical evaluation of the integral yields:
\begin{equation}
\label{eq:app_kernel_final_value}
K_\theta(0,0;0) \approx 1 - \frac{\im U}{4} (96.43\im) = 1 + 24.1 U.
\end{equation}
This result is positive definite for all repulsive interactions $U>0$. This implies that the static orbital susceptibility is finite and non-divergent. 

\begin{figure}[!htp] 
    \centering
    \includegraphics[width=0.8\columnwidth]{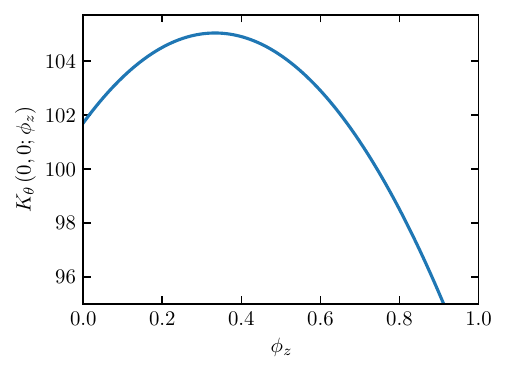}
    \caption{The orbital stiffness kernel $K_\theta(0,0;\phi_z)$ as a function of the staggered magnetization field amplitude $\phi_z$. The kernel remains positive definite across the entire range from the semimetallic phase ($\phi_z=0$) to the gapped AFM phase ($\phi_z > 0$). A positive value corresponds to a finite mass for the orbital fluctuations, indicating that the orbital excitations are gapped and the system is stable against orbital ordering.}
    \label{fig:orbital_gap}
\end{figure}

For the AFM phase where $\phi_z > 0$, the structure of the kernel $K_\theta(0,0;\phi_z)$ remains formally identical. We evaluate this kernel numerically as a function of the magnetic order parameter $\phi_z$. As shown in Fig.~\ref{fig:orbital_gap}, the kernel $K_\theta$ remains strictly positive throughout the transition from the semimetal to the AFM insulator. Since the condition for a gapless mode (an instability) is $K_\theta(0, 0) \leq 0$, our result $K_\theta > 0$ confirms that the orbital excitations remain massive (gapped) across the phase diagram. This provides robust field-theoretic evidence that the system is stable against spontaneous orbital ordering, justifying our treatment of the orbital degree of freedom as an unordered channel.

\section{Numerical Evidence for Double-Exchange Mechanism}
\label{app:double_exchange_evidence}
To further substantiate the double-exchange interpretation near half filling, we compare the zero-temperature thermodynamics of the constrained homogeneous phases. At fixed interaction $U=8t$, we compute the internal energy density $\mathcal{E}(n)$ for both the FM and AFM states as a function of filling $n$ by constraining the system to each order. In the crossing region, the mean-field ground state at $T=0$ minimizes the total energy via a common-tangent (Maxwell) construction, yielding a density-discontinuous interval and hence phase separation. Thermodynamically, this is equivalent to minimizing the grand potential $\Omega(\mu) = e - \mu n$. Coexistence occurs when the two phases share the same chemical potential $\mu = \partial e / \partial n$ and the same grand potential $\Omega_{FM}(\mu) = \Omega_{AFM}(\mu)$. For any average filling between the two tangent endpoints, the lower energy is given by a linear mixture of the two phases. This behavior is standard in double-exchange-driven systems and is widely discussed in phase-separation scenarios for colossal-magnetoresistive materials~\cite{yunoki_phase_1998,tokura_orbital_2000}. At fixed filling $n=1.85$, the real-space mixed-state calculation yields a minimum at intermediate FM fraction, consistent with coexistence near the first-order boundary.
\begin{figure}[!htp]
    \centering
    \includegraphics[width=\columnwidth]{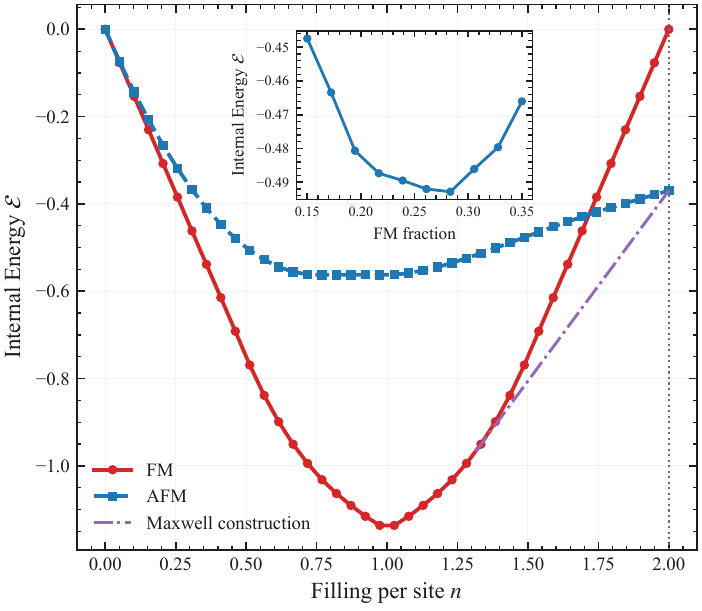}
    \caption{Additional numerical support for phase competition near half filling at $U=8t$. Energy per site versus filling $n$ for homogeneous FM and AFM solutions, with Maxwell construction indicating a density-discontinuous region consistent with phase separation. Inset: energy per site versus FM fraction at fixed filling $n=1.85$.}
    \label{fig:app_double_exchange_evidence}
\end{figure}

\section{Mean-Field Calculation Details}
\label{app:mean_field_details}
The ground-state phase diagrams were determined using a self-consistent mean-field theory. For this calculation, the on-site interaction Hamiltonian $H_{\text{int}}$ is expressed using the orbital pseudo-spin operators $T^j_\sigma = \frac{1}{2} \sum_{\mu,\mu'} c_{\mu\sigma}^\dagger \tau^j_{\mu\mu'} c_{\mu'\sigma}$:
\begin{equation}
\label{eq:interaction_T_operators_appendix}
H_{\text{int}} = \frac{U}{2} T^0_\uparrow T^0_\downarrow + \frac{U}{4} T^x_\uparrow T^x_\downarrow + \frac{U}{4} T^z_\uparrow T^z_\downarrow.
\end{equation}
Note that this expression is exactly equivalent to the form presented in Eq.~\ref{eq:interaction_S_operators} in the main text.
Applying this decoupling to Eq.~(\ref{eq:interaction_T_operators_appendix}) results in a quadratic mean-field Hamiltonian $H_{MF} = H_t + \sum_i H_{int,i}^{MF}$. The local interaction term, written explicitly in terms of the order parameters $t^j_{i,\sigma} = \langle T^j_{i,\sigma} \rangle$, is:
\begin{align}
\label{eq:app_hamiltonian_mf_interaction_simplified}
H_{int,i}^{MF} = &\frac{U}{2} \left(t^0_{i,\downarrow} T^0_{i,\uparrow} + t^0_{i,\uparrow} T^0_{i,\downarrow}\right) \nonumber \\
&+ \frac{U}{4} \left(t^x_{i,\downarrow} T^x_{i,\uparrow} + t^x_{i,\uparrow} T^x_{i,\downarrow}\right) \nonumber \\
&+ \frac{U}{4} \left(t^z_{i,\downarrow} T^z_{i,\uparrow} + t^z_{i,\uparrow} T^z_{i,\downarrow}\right) - E_{\text{sub},i},
\end{align}
where $E_{\text{sub},i} = \frac{U}{2} t^0_{i,\uparrow} t^0_{i,\downarrow} + \frac{U}{4} t^x_{i,\uparrow} t^x_{i,\downarrow} + \frac{U}{4} t^z_{i,\uparrow} t^z_{i,\downarrow}$ is the standard double-counting correction. The full Hamiltonian $H_{MF}$ is then solved self-consistently by diagonalizing it to find the new ground state expectation values $\langle T^j_{i,\sigma} \rangle$ until convergence is reached.

To ensure the robustness of our findings, we employed both momentum-space and real-space methods. First, we performed an unbiased search for the dominant ordering patterns in momentum space. The site-dependent order parameters were Fourier transformed, $t^j_{\alpha,\sigma}(\vec{q}) = \frac{1}{\sqrt{N}} \sum_{i \in \alpha} e^{-i\vec{q}\cdot\vec{r}_i} t^j_{i,\sigma}$, where $\alpha \in \{A,B\}$ is the sublattice index. In our initial calculation, the Fourier components $t^j_{\alpha,\sigma}(\vec{q})$ were treated as free parameters for all wavevectors $\vec{q}$ in the Brillouin zone. This comprehensive search consistently found that the self-consistent solution minimized the energy only when $t^j_{\alpha,\sigma}(\vec{q})$ was non-zero for the uniform mode ($\vec{q}=0$, ferromagnetic) or the staggered mode ($\vec{q}=K$, antiferromagnetic). To corroborate this, a complementary real-space calculation was performed on a finite diamond-shaped cluster (characteristic length of 21 sites), which, after iterating from numerous random initial states, confirmed the stability of only these FM and AFM phases. Then, having established the nature of the competing ground states, we mapped out the detailed phase diagrams using a simplified and computationally efficient model, where the self-consistent calculation was restricted to the subspace of order parameters with only $\vec{q}=0$ or $\vec{q}=K$ components, which yielded the magnetic phase diagram presented in Fig.~\ref{fig:main_phase_diagram}.

\section{Derivation of the Antiferromagnetic Gap Equation}
\label{app:gap_equation}

The onset of the staggered antiferromagnetic (AFM) phase at half-filling is described by a gap equation for the order parameter $m_s$. With the order parameter defined as in Eq.~(\ref{eq:staggered_magnetization_order_parameter}), the mean-field Hamiltonian can be written as 

\begin{equation}
H_{\mathrm{MF}}(\vec{k})=
H_t(\vec{k})+Um_s\,\sigma^z_{\mathrm{spin}}\otimes\rho^z_{\mathrm{sub}},
\end{equation}
where $\rho^z_{\mathrm{sub}}=\pm 1$ on the $A/B$ sublattices. The quasiparticle energies are
\begin{equation}
\begin{aligned}
E_D(\vec{k})&=\sqrt{\epsilon(\vec{k})^2+(Um_s)^2}, \\
E_f&=\sqrt{(3/2)^2+(Um_s)^2},
\end{aligned}
\end{equation}
for Dirac and flat sectors, respectively. At $T=0$, minimizing the mean-field free energy
\begin{equation}
f(m_s)=Um_s^2-\!\!\int_{\mathrm{BZ}}\!\!\frac{d^2k}{\Omega_{\mathrm{BZ}}}\,E_D(\vec{k})-E_f
\end{equation}
with respect to $m_s$ gives
\begin{equation}
\label{eq:app_gap_equation}
      \frac{4}{U} = \int_{\text{BZ}} \mathrm{d}^2k  \frac{3\sqrt{3}}{8\pi^2} \frac{1}{\sqrt{\epsilon(\vec{k})^2 + (Um_s)^2}}
+ \frac{1}{\sqrt{\frac{9}{4} + (Um_s)^2}},
\end{equation}
This is the self-consistent AFM gap equation used in the main text. It can be rewritten in terms of the density of states of the Dirac bands, $g(\epsilon)$, as
\begin{equation*}
    \frac{4}{U} = \frac{1}{\sqrt{\frac{9}{4} + (Um_s)^2}} + \int d\epsilon \, \frac{g(\epsilon)}{\sqrt{\epsilon^2 + (Um_s)^2}}.
\end{equation*}
To determine the critical interaction strength $U_c$ for the onset of staggered magnetization, we analyze the gap equation in the limit $m_s \to 0$:
\begin{equation}
\label{eq:app_analytical_Uc}
\frac{4}{U_c} = \frac{3\sqrt{3}}{4\pi^2} \int_{\text{BZ}} \frac{\mathrm{d}^2k}{|\epsilon(\vec{k})|} + \frac{2}{3},
\end{equation}
where the integral accounts for the Dirac bands and the constant term arises from the flat bands. A precise numerical evaluation of the integral yields the critical interaction strength $U_c \approx 3.253t$. This result is independently confirmed by our full self-consistent mean-field calculations.

\end{document}